\documentclass[conference]{IEEEtran}
\usepackage{blindtext, graphicx}
\usepackage{csquotes}

\ifCLASSINFOpdf
\else
\fi
\begin{document}
\title{Performance of Distributed File Systems on Cloud Computing Environment: An Evaluation for Small-File Problem}

\author{
\IEEEauthorblockN{Thanh Duong}
\IEEEauthorblockA{John von Neumann Institute\\
Vietnam National University at HCMC\\
E-mail: thanh.duong@jvn.edu.vn}
\and
\IEEEauthorblockN{Quoc Luu}
\IEEEauthorblockA{John von Neumann Institute\\
Vietnam National University at HCMC\\
E-mail: quoc.luu2015@qcf.jvn.edu.vn}

\and
\IEEEauthorblockN{Hung Nguyen}
\IEEEauthorblockA{Faculty of Economics and Management\\
Thai Binh Duong University\\
E-mail: hung.nb@tbd.edu.vn}
}
\IEEEoverridecommandlockouts
\IEEEpubid{\makebox[\columnwidth]{978-1-5090-4134-3/16/\$31.00~\copyright~2016~IEEE \hfill} 
\hspace{\columnsep}\makebox[\columnwidth]{ }}

%Telephone: (800) 555--1212, Fax: (888) 555--1212}


\maketitle

\begin{abstract}
%\boldmath
% \blindtext[1]


Various performance characteristics of distributed file systems have been well studied. However, the performance efficiency of distributed file systems on small-file problems with complex machine learning algorithms scenarios is not well addressed. In addition, demands for unified storage of big data processing and high-performance computing have been crucial. Hence, developing a solution combining high-performance computing and big data with shared storage is very important. This paper focuses on the performance efficiency of distributed file systems with small-file datasets. We propose an architecture combining both high-performance computing and big data with shared storage and perform a series of experiments to investigate the performance of these distributed file systems. The result of the experiments confirms the applicability of the proposed architecture in terms of complex machine learning algorithms.

\end{abstract}

% \fi
\IEEEpeerreviewmaketitle

\section{Introduction}
% \blindtext

From creating big models in machine learning to answering big questions in data processing, big data has been widely adopted and growing as a promising technology for enterprises to develop and improve their businesses efficiently. In research disciplines, along with high-performance computing, big data also plays a critical role in data processing \cite{Fox:fo, Moise:2016ff, Zhao:bd}. However, high-performance computing mainly uses parallel distributed file systems for I/O and very limited local storage, whereas big data processing relies on Hadoop distributed file system (HDFS) and local storage \cite{Bhat:2016eu}. In addition to storage issues, the output of scientific applications is often very large and would take a long time to transfer to external big data processing clusters (i.e. Hadoop system). Furthermore, the principle \textquote{write-once, read-many} of HDFS design does not meet high-performance computing requirements \cite{Xuan:2015hy}. Thus, it requires an approach to provide a unified storage solution.

Hadoop distributed file system (HDFS) is an open-source distributed storage platform that implements a data processing framework called MapReduce on top of it. Many companies (e.g., AOL, Yahoo!, Facebook) use HDFS to manage and store petabyte or exabyte-scale enterprise data on clusters due to its reliability and scalability \cite{Bhat:2016eu, Rao:2012td}. A fundamental philosophy of Hadoop system design is that \textquote{moving computation is cheaper than moving data} to leverage local storage (direct attached storage - DAS) and reduce network transfer \cite{Borthakur:2007tw}. In other words, it means that Hadoop achieves more efficiency locally executing computations on nodes storing the data involved rather than moving it to other compute nodes across the network \cite{Rutman:2011tp}. Hadoop can run on various hardware, especially commodity hardware. For instance, instead of using RAID to achieve fault tolerance, Hadoop replicates data across machines in a large cluster \cite{Borthakur:2007tw}. In addition to Hadoop system, Spark is a data processing framework running on top of HDFS and YARN to leverage data locality like MapReduce of Hadoop system \cite{Zaharia:2010wg}. As an alternative to MapReduce framework, Spark has been widely adopted for big data processing due to its speed, easiness, and complement to MapReduce framework limitations \cite{Srinivasa:2015tk}. Moreover, beyond HDFS, Spark can also interface with a wide variety of distributed file systems (e.g., MapR File System, Cassandra, OpenStack Swift, Amazon S3, Chukwa, and Lustre) and local file system \cite{Chaimov:2016go,Diaz:2016ff,Zaharia:2010wg}.

Lustre is a massively global and parallel distributed file system for high-performance computing \cite{Raj:2015vu}. It is among the most commonly deployed on supercomputing clusters \cite{Anonymous:GomU6HJ5}. Lustre provides POSIX-compliant with more than a terabyte per second (TB/s) of aggregate I/O throughput \cite{Anonymous:5056JV_l}. With InfiniBand, it will be able to directly transfer data from virtual memory on one node to virtual memory on another node for improvement of throughput and CPU usage  \cite{Yu:2006wp}. However, for data-intensive jobs with large datasets, researchers argue that network transfers should be a potential bottleneck on non-HDFS systems \cite{Chaimov:2016go,Dean:2008fi,Zaharia:2010jf}. In contrast, Ananthanarayanan et al. suggest that data locality becomes irrelevant when networking technology is improving faster than the disk speeds \cite{Ananthanarayanan:2011vp}. Furthermore, Xuan et al. suggest that the design of high-performance computing architecture provides high storage capacity with low-cost data fault tolerance \cite{Xuan:2015hy}.

Since there are many challenges of high-performance computing and big data processing combination, it is useful to re-examine Spark with both data-intensive and compute-intensive under HDFS and non-HDFS. Furthermore, the small-file problem is an important criterion for examining the efficiency of distributed file systems because the storage of numerous small files reduces the efficiency of clusters. As a result, the paper aims to study the applicability of a unified architecture combining high-performance computing and big data processing. In this paper, we expect that the performance of both Lustre and HDFS file systems will be able to achieve equivalently for big data analytics without leveraging data locality.

\section{Related Work}
Performance evaluation of distributed file systems with big data processing frameworks has been the subject of numerous recent studies \cite{Moise:2016ff, Rutman:2011tp, WasiurRahman:2015cu, Xuan:2015hy, Zhao:2015ce}. In MapReduce applications with a local disk for the shuffle phase, Zhao et al. found that the performance of the I/O throughput of Hadoop-Lustre is nearly equal to Hadoop-HDFS \cite{Zhao:2015ce}. Especially, TestDFSIO benchmark shows that the read operation of Hadoop-Lustre always achieves higher performance than Hadoop-HDFS. Rutman \cite{Rutman:2011tp} suggests that using Lustre as an HDFS file system replacement on an \textquote{expensive} Hadoop compute clusters in HPC environments can significantly improve overall \textquote{system performance and cluster efficiency or potentially offer a lower overall system cost}. Furthermore, researchers use Tachyon as an in-memory file system with OrangeFS distributed filesystem to deploy data-intensive frameworks in HPC clusters \cite{Xuan:2015hy}. The result of the theoretical model indicates an improvement in read-throughput. Also, it will be able to scale up with the number of compute nodes for Hadoop.

Leveraging RDMA of Infiniband on HPC clusters to increase horizontal data movement has been the target of many studies \cite{Bhat:2016eu,  Lu:2014kx, Shankar:2016ef}. Bhat et al. show that RDMA-based HDFS plug-in significantly outperforms original Hadoop distributions (i.e., Cloudera, Hortonworks). The approach, which accesses the native communication library via JNI, can be deployed on many Hadoop distributions with many different versions to leverage the high-speed interconnect of HPC clusters (i.e., InfiniBand). Moreover, many attempts (e.g., Hadoop Adapter for Lustre from Intel and Apache Hadoop on Lustre Connector from Seagate) to optimize data movements in the shuffle phase of Hadoop have been carried out \cite{IntelCorporation:H4v93YQU, IntelCorporation:CzNi7b54, Anonymous:X0T8arn9}. In contrast to RDMA-based HDFS plugins, these Hadoop plug-ins for Lustre propose an approach called \textquote{shuffle with zero-copy}. In other words, this approach, which allows any data to be available to all compute nodes at any time, uses Lustre as a centralized storage. Hence, it eliminates transfers during the MapReduce shuffle phase to achieve higher job throughput \cite{Anonymous:X0T8arn9}.

\section{Statement of Problems}
% \blindtext
\begin{figure*}[t]
    \centering
    \includegraphics[scale=0.248]{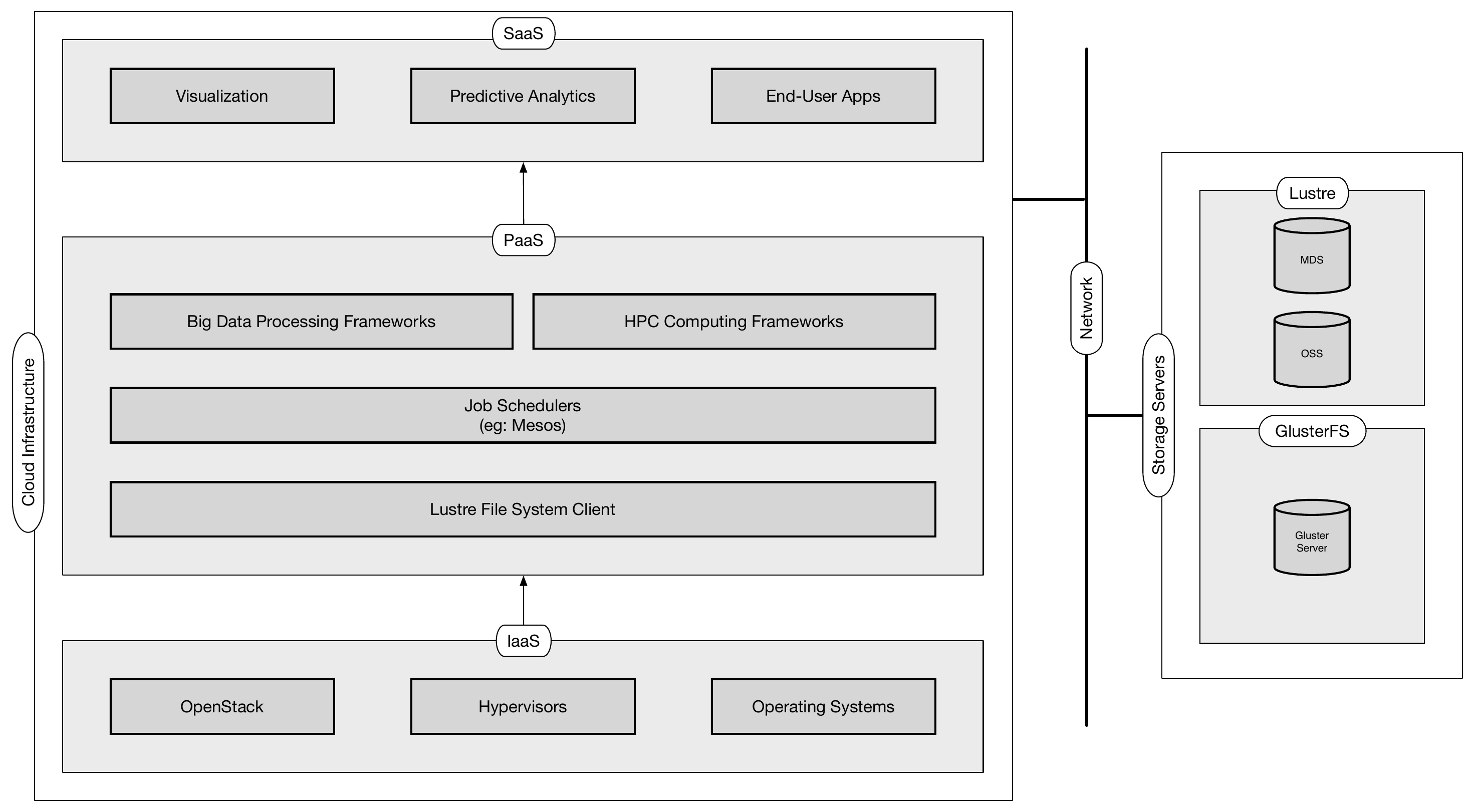}
	\par \begin{flushright}
    \end{flushright}
    \caption{Proposed Architecture for high-performance Computing and Big Data Processing}
    \label{figure:ProposedArchitecture}
\end{figure*}
\subsection{The Movement of Data in Spark}
% \blindtext

Data movement is one of the key factors affecting the performance of a Spark application in any distributed system. Essentially, there are two types of data movement. They are vertical data movement and horizontal data movement \cite{Chaimov:2016go} as shown in our proposed architecture in Figure \ref{figure:ProposedArchitecture}.

\begin{figure}
    \centering
    \includegraphics[scale=0.23]{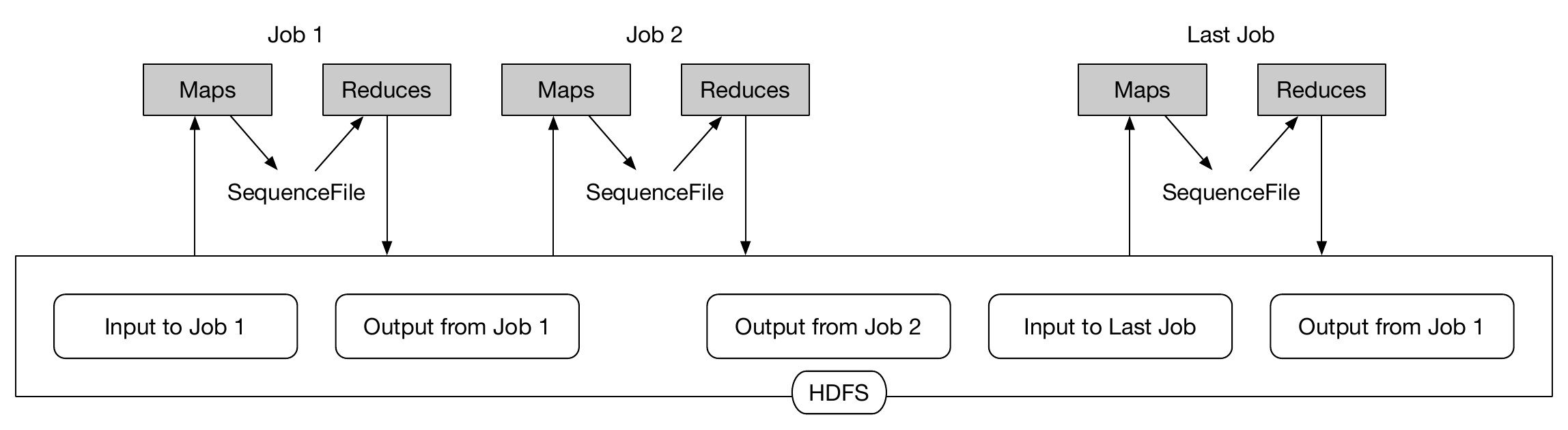}
	\par \begin{flushright}
    \textit{Source: Adopted from \cite{McDonald:QG8JVfyi}}
    \end{flushright}
    \caption{Iteration jobs in MapReduce}
    \label{figure:IterationMR}
\end{figure}

The first type refers to \textquote{the movement through the entire memory hierarchy, including persistent storage}. Block of data from various input sources will be moving into memory to create resilient distributed datasets (RDDs) for processing \cite{Hamstra:2015we}. By default, RDDs are lazy and ephemeral; they are recomputed after every action is run. However, with the LRU (Least Recently Used) eviction rules, Spark allows developers to persist (i.e., \textit{save} and \textit{cache}) data in a number of different places (e.g., in memory and on disk) for reuse to minimize vertical movement for RDDs \cite{Chaimov:2016go,Hamstra:2015we,Zaharia:2010wg}. Persisting data in Spark is more useful for iterative algorithms, especially in machine learning (e.g., Gradient-Descent for parameter optimization) \cite{Srinivasa:2015tk}. Conversely, instead of utilizing memory for storing intermediate working sets, Hadoop MapReduce framework loads data from disk on each iteration \cite{White:2012vl}. As Figure \ref{figure:IterationMR} suggests, due to the requirement of many I/O activities, Hadoop MapReduce framework leads to the increment of bandwidth usage. In other words, with iteration algorithms, data locality is not as important in Spark as in MapReduce. Despite the advantage of persisting data, the capacity of memory is limited. The block manager of Spark runtime can drop or move cached data across the memory hierarchy. Additional vertical data movement for re-computation will be needed. Spark is designed with a smart algorithm enough to be able to deal with the trade-off between storing RDDs, speed of accessing it, probability of losing it, and cost of re-computing \cite{Srinivasa:2015tk}. In addition to reducing vertical data movement for big data processing frameworks, project Tachyon helps these frameworks decrease the re-computation and transfer time by using data lineage and fast memory \cite{Li:2013vd}.

\begin{figure*}
    \centering
    \includegraphics[scale=0.38]{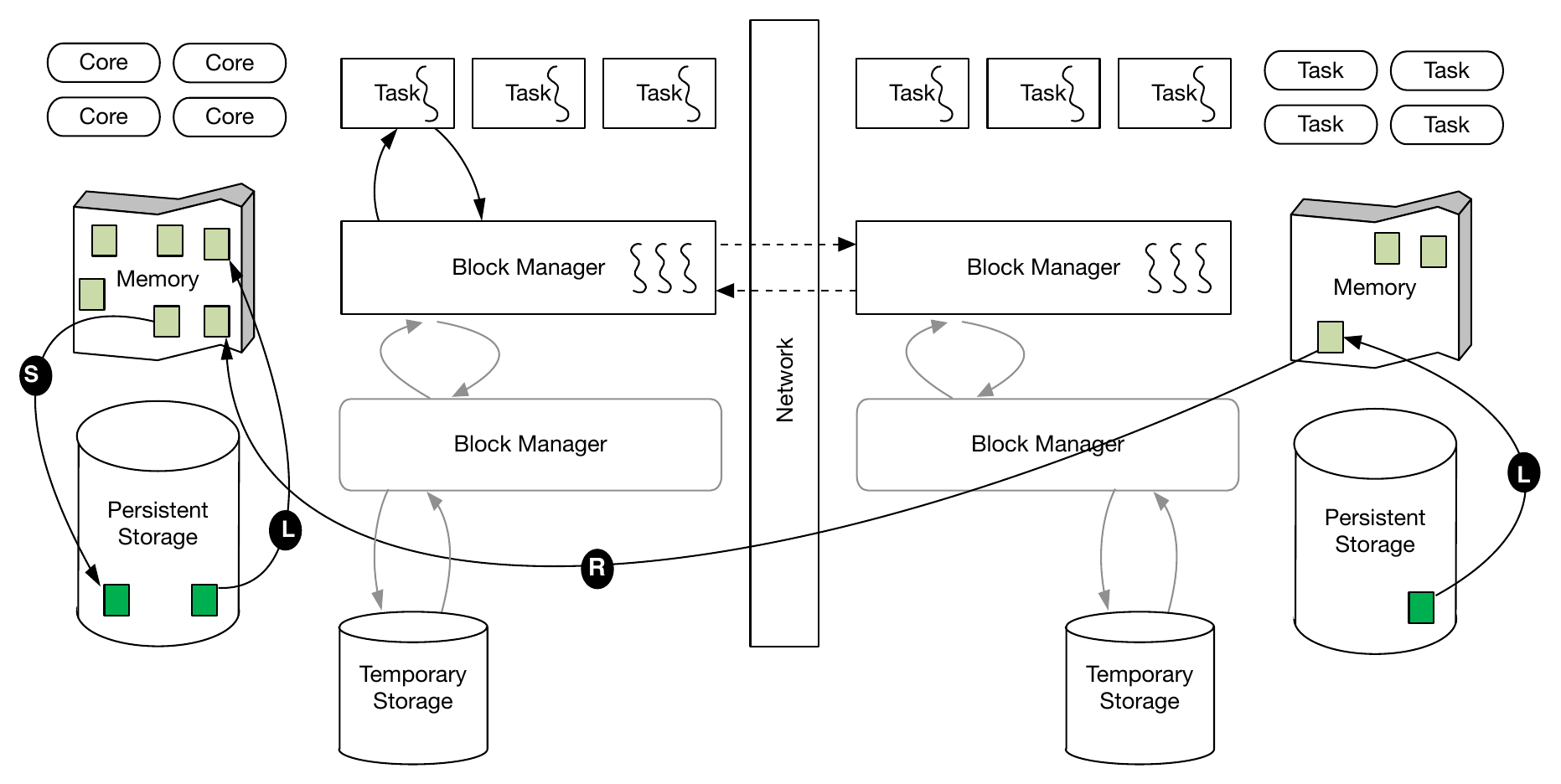}
	\par \begin{flushright}
    \textit{Source: Adopted from \cite{Chaimov:2016go}}
    \end{flushright}
    \caption{Data movement in Spark and the interaction with the memory hierarchy}
    \label{figure:DataMovementSpark}
\end{figure*}

The next data movement refers to \textquote{the shuffle communication phase between compute nodes} \cite{Chaimov:2016go}. The shuffle manager is responsible for managing this data movement, along with the block manager. For scratch storage during shuffle operations, Spark uses a local disk to handle the whole process \cite{Hamstra:2015we}. There are many ways to optimize the performance of horizontal data movement. The first approach uses built-in compression techniques in Spark to compress shuffle outputs, and data is spilled during shuffles. The next approach is leveraging RDMA interconnection in high-performance computing clusters to accelerate shuffle data \cite{Lu:2014kx}. However, the level of parallelism is highly dependent on every practical problem. Hence, these techniques do not always intensely impact the overall performance of a Spark application due to the dominance of vertical data movement \cite{Chaimov:2016go, Sparks:2014tu}.

\subsection{The Movement of Data in System Architectures}
% \blindtext

From the beginning, Hadoop system, which is based on Google File System (GFS) and Google MapReduce framework, is designed to work with commodity hardware for data-intensive applications with local storage to reduce the communication cost and replication data to achieve fault tolerance \cite{Ghemawat:2003dy, Dean:2008fi,  White:2012vl}. The Hadoop system provides a linear scale by adding commodity servers to increase computation capacity and storage capacity \cite{Shvachko:2010gw}. The system will likely achieve petascale \cite{Fasale:2015ue, White:2012vl}. By default, HDFS splits a dataset into 128-MB blocks. The block size can be changed for the cluster or any specified file when it is created \cite{White:2012vl}. Typically, datasets stored in HDFS can be varied from gigabytes to terabytes in size \cite{Khan:2015ws}. Like other distributed file systems, HDFS stores file system metadata and application data separately \cite{Shvachko:2010gw, Wang:2009vg}. HDFS assigns NameNode servers to store metadata and DataNodes servers to store application data. The HDFS Federation feature allows multiple NameNode servers to operate simultaneously and actively to avoid bottlenecks. Every communication protocol of HDFS is layered on top of the TCP/IP protocol \cite{Borthakur:2007tw}. In the client node, the HDFS client determines the location of blocks using remote procedure calls (RPC). However, HDFS is unable to be mounted and does not fully support POSIX-compliant because the file system is optimized for specific workloads to help targeted applications of HDFS achieve higher performance by eliminating many hard requirements of POSIX-compliant file system \cite{Borthakur:2007tw, Moise:2010uh}. In other words, this traded feature helps HDFS to increase data throughput rates, especially the cost of support for non-POSIX operations such as Append \cite{Borthakur:2007tw}. Moreover, as a complement to the principle \textquote{write-once, read-many}, HDFS is designed for immutable files with \textquote{single-writer, multiple-reader} model \cite{Shvachko:2010gw}. More importantly, in the Hadoop system, one of the most important components is the resource-management platform YARN (Yet Another Resource Negotiator) \cite{Vavilapalli:2013eu}. Before YARN, Hadoop restricted the processing model of batch-oriented MapReduce jobs with resource management infrastructure. Vavilapalli at el. argue that this inadequacy forces developers to abuse the MapReduce programming model \cite{Vavilapalli:2013eu}. In the next generation of Hadoop, MapReduce is just one of the frameworks running on top of YARN. Regarding data locality optimization, YARN assigns JobTracker to place every task close to their input data in HDFS. Hence, as mentioned before, in the world of Hadoop, a compute node should be the same as a data node to leverage this design. Despite many optimizations in architecture design, there are some drawbacks of Hadoop HDFS. Firstly, Li et al. state that the I/O throughput of HDFS is limited due to data replication for fault-tolerance \cite{Li:2013vd}. To overcome this limitation, using RAID configurations is still useful \cite{Xu:2012uz}. Furthermore, with a RAID 6, the capacity and performance of storage devices are more efficiently utilized. Secondly, in terms of data storage capacity and deployment cost, using a separate sub-cluster for storage (high-performance computing architecture) is more efficient than deploying DAS on compute nodes \cite{Xuan:2015hy}. Finally, as suggested by Ananthanarayanan et al., the \textquote{moving computation is cheaper than moving data} principle does not reflect the recent change of technology when bandwidth growth and cost declines \cite{Kipp:bUo_VuOm, Lago:2015bp, Law:2012wa, Morgan:2015wp}. Also, by leveraging fast memory, project Tachyon indicates that for MapReduce applications, locality data should be irrelevant on non-HDFS systems \cite{Xuan:2015hy, Li:2013vd}.

Lustre is a storage architecture for clusters only supported on the Linux operating system \cite{IntelCorporation:ONhglhQx}. In contrast to HDFS of Hadoop system, Lustre uses its own file system (an improved version of the ext4 journaling file system) as a back-end to store data and metadata and supports POSIX-compliant file system interface allowing mounting from clients \cite{Meshram:2011up, Wang:2009vg}. Moreover, Lustre also supports ZFS as a back-end to use many advantages of this file system \cite{Grandinetti:2015wq}. For instance, ZFS file system allows storing over 1.5 million files in a single directory, online data integrity checking, and error handling, high data compression ratio without reducing performance \cite{Stearman:2015ub}. By aggregating the storage capacity and throughput of many servers, Lustre storage cluster permits clients to nearly fully utilize the I/O throughput of interconnect devices with low latency. Hence, Lustre storage architecture is adopted for various purposes, from home users to high-performance computing clusters \cite{Anonymous:2014uw}. It is best known for serving many of the largest supercomputers around the globe, with thousands of client systems, petabytes (PB) of storage, and over a terabyte per second (TB/sec) of I/O throughput \cite{ Anonymous:GomU6HJ5, Anonymous:5056JV_l, IntelCorporation:ONhglhQx}. In terms of scalability, like HDFS of Hadoop system, Luster allows adding servers dynamically to increase both throughput and capacity of storage. Lustre also supports RDMA for InfiniBand to achieve higher I/O throughput, network transport efficiency and reduced CPU usage. Basically, a Lustre file system is made up of three components: MDS, OSS, and client \cite{IntelCorporation:ONhglhQx}. This first component is one or more \textit{Metadata Server} (MDSs - server nodes) with one or more \textit{Metadata Target} (MDTs - persistent storage like SSD, HDD) to store metadata (such as filenames, directories, permissions, and file layout). The MDS does not involve a file I/O process once the client obtains stripped information from the file. Instead, the client directly interacts with OSSs of the Luster file system to perform I/O operations. The second component is one or more \textit{Object Storage Server} (OSSs - server nodes) with one or more\textit{ Object Storage Target} (OSTs - persistent storage like SSD, HDD) to store one or more objects (depend upon stripe configuration and size of file), each object on a separate OST across the Lustre file system. Along with other high-performance computing applications, in Spark applications, the throughput of vertical data movement on the Lustre file system is the aggregated throughput provided by each OST. Likewise, the capacity of the Lustre file system is the total capacity provided by each OST. The final component is one or more clients to mount the Lustre file system as an accessible file system. In contrast to HDFS, Lustre file system is carefully optimized for \textquote{many-write-many-read} principle, which is a critical requirement for scientific applications \cite{Zhao:2015fz}. In other words, Luster fully supports concurrent and coherent read and write access to the files in the file system. Lustre does not support data redundancy. To achieve failover, it must rely on backup techniques like RAID. Like Hadoop HDFS, despite many optimizations in architecture design, there are some drawbacks of Lustre file system, which will be covered in the next section.

\begin{figure}
    \centering
    \includegraphics[scale=0.35]{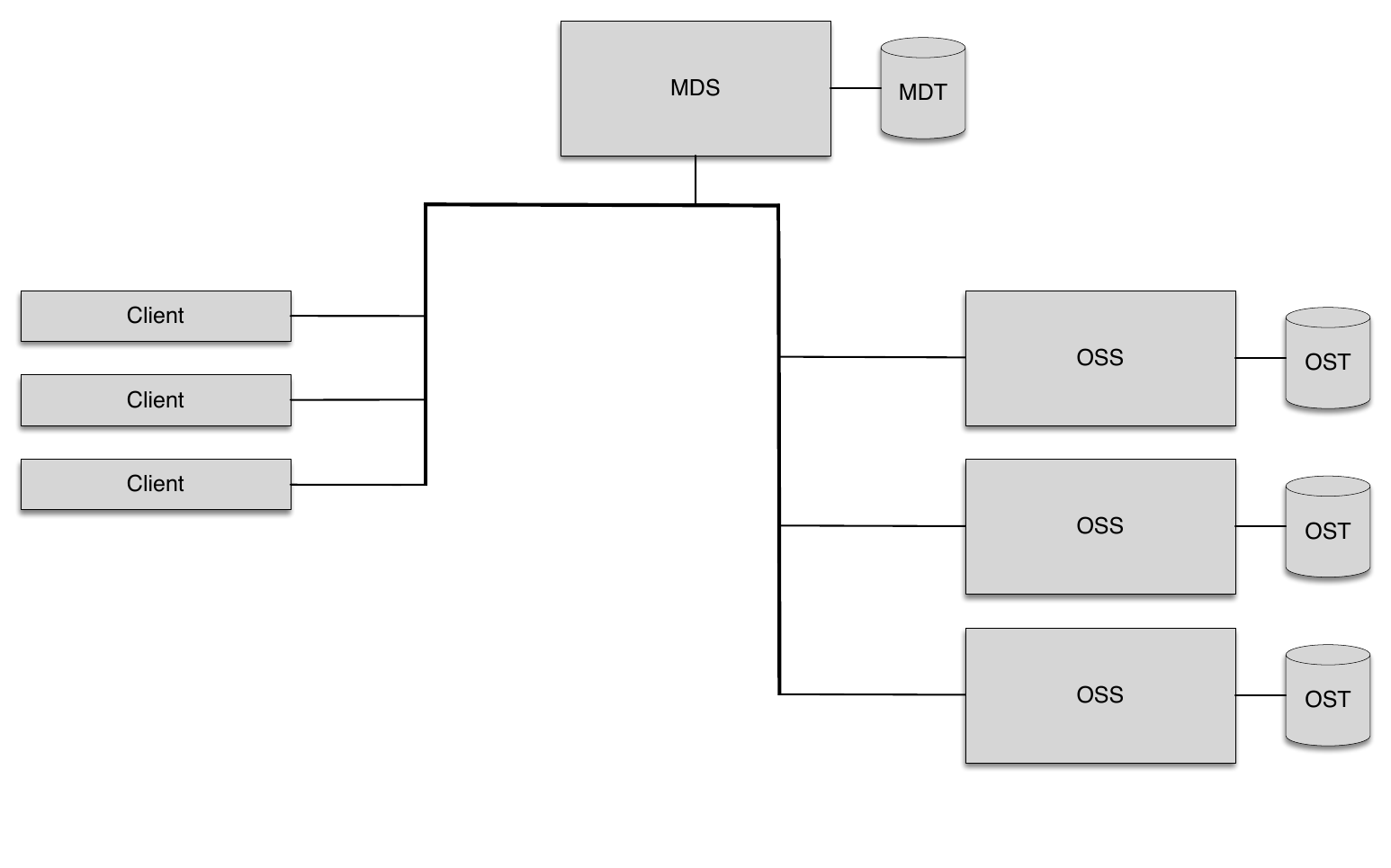}
	\par \begin{flushright}
    \textit{Source: Adopted from \cite{IntelCorporation:ONhglhQx}}
    \end{flushright}
    \caption{Example of Lustre cluster layout}
    \label{figure:LustreArch}
\end{figure}

\subsection{The Small-File Problem of Distributed File System}
% \blindtext

Every distributed file system is designed with the assumption that the file size is very large. Thus, its metadata management becomes bottle-necked because of handling large numbers of small-file requests from clients \cite{Shvachko:2010wh}. 

Due to the nature of both scientific applications and many Internet-oriented applications, dealing with a large number of small files is unavoidable \cite{Wang:2015fh, Zhao:2015fz}. Accordingly, \textquote{the percentage of files smaller than 1 KB is 49.3\%, and only 0.8\% files are larger than 10 MB} \cite{Wang:2015fh}. In HDFS, namenode keeps metadata of the filesystem in memory, and small files dramatically increase the amount of memory for the metadata on the namenode \cite{White:2012vl}. The number of map tasks is equal to the amount of files. Hence, it leads to an overhead parallelism level, which is tens or hundreds of times slower than the equivalent job with a single input file \cite{White:UVECqG-G}. Moreover, it takes about 7.7 hours to store 550000 small files, whose sizes range from 1KB to 10KB, into HDFS while in ext3 local file system, it is only 660 seconds \cite{Liu:2009jq}. Likewise, many researchers argue that small files significantly impact the overall performance of Lustre file system \cite{Meshram:2011up, Pershin:ZJwZW-GL}. The number of operational MDS is restricted to one; the other MDS becomes operational if the first MDS crashes. Hence, it remains a potential bottleneck as the number of clients and/or files increases significantly. However, Lustre Distributed Namespace (DNE) projects allow multiple pairs of active/active metadata servers to improve Lustre metadata performance and scalability \cite{Simmons:qjlt2KK3}. In addition, for every client request, two RPCs are generated at least \cite{Meshram:2011up}. It is very costly to request many small files concurrently. Moreover, like HDFS, in Lustre, the number of map tasks equals the number of files.

Hence, dealing with small files is a serious problem. It requires many advanced techniques, including a back-end modified layer of Lustre and/or powerful hardware combination to overcome the issue \cite{Xuan:2015hy, Ihara:Xcb_qNAB, Li:2013vd, Meshram:2011up, Moise:2016ff, Simmons:qjlt2KK3}. As mentioned in this paper, we focus on the performance efficiency of distributed file systems on small files issues with complex machine learning algorithms scenarios. Hence, we do not include these suggestions in the proposed architecture.

\section{Empirical Results}
% \blindtext

\subsection{Experimental Setup}
% \blindtext

The experiment was performed on two clusters, which ran on the top of KVM (Kernel-based Virtual Machine). In Lustre cluster, we deployed Lustre 2.7 on a 2-core, 2GB of RAM with a 100-GB MDT for the MDS and 2-core, 2GB of RAM with a 100-GB OST for each OSS. However, the allocated resource for Lustre cluster storage does not fully utilize the read-only caching of data feature in the OSS to enhance performance. In Hadoop cluster, we deployed Cloudera Distribution Hadoop 5.7.1 (including Spark 1.6). The Cloudera master node had a 6-core with 24GB of RAM and a 6-core with 12GB of RAM for each worker. Also, the Hadoop cluster could mount Lustre file system from the Lustre cluster. Each physical node uses a 1-Gigabit TCP/IP Ethernet.

On our benchmark, we use Kaggle Display Advertising Challenge Dataset from Criteo with functions (i.e., \textit{LogisticRegressionWithSGD}, \textit{LabeledPoint} and \textit{SparseVector}) from Spark MLlib \cite{Criteo:6gTDLCJB}. We split the training dataset from Criteo into multiparts, and each part contains 460 lines. We use 10000-file (1.1 GB in total), 20000-file (2.2GB in total), and 30000-file (3.3GB in total) for the benchmark. Moreover, we use Spark Standalone for the data source of jobs from Lustre, and we use Spark with YARN scheduler for the data source of jobs from HDFS.

\subsection{Result Analysis}
% \blindtext

\begin{figure}[t]
    \centering
    \includegraphics[scale=0.53]{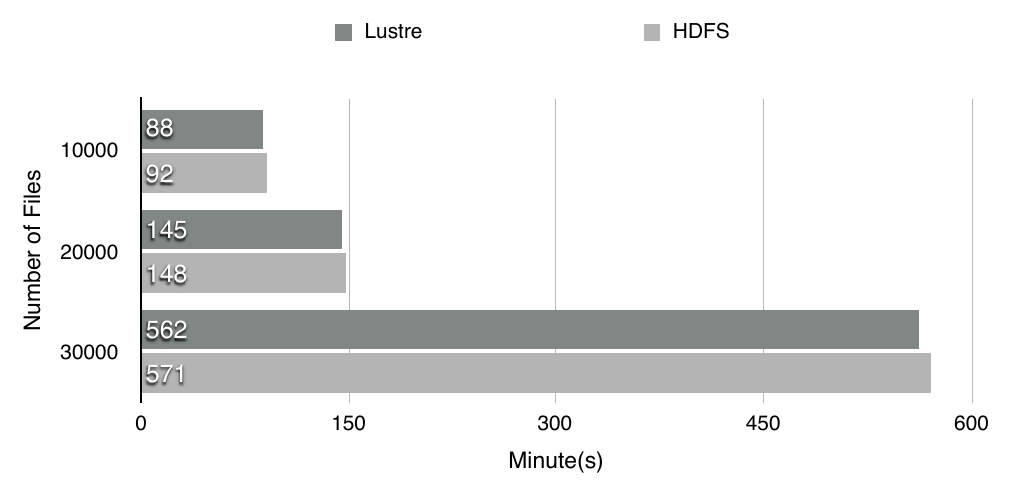}
	\par \begin{flushright}
    \end{flushright}
    \caption{Execution time of jobs on Lustre and HDFS file system with different amounts of files}
    \label{figure:BM}
\end{figure}

As Figure \ref{figure:BM} suggests, the execution time of jobs on both Lustre and HDFS file systems increases quickly as the amount of data for processing becomes larger. Especially when dealing with 30000 files, the running time on Lustre and HDFS file systems dramatically increases.

The execution time of jobs in Spark with Lustre slightly outperforms Spark with HDFS. However, it is substantial. Moreover, the gap increases when the number of small files is larger. In other words, regarding the small file problem, Lustre is appropriate for big data storage.

As a result, as Figure \ref{figure:BM} indicates, a unified storage solution for both high-performance computing and big data processing is achievable by using Luster distributed file system for both jobs.

\section{Conclusion}
% \blindtext
In this paper, we design and propose a unified platform for high-performance computing and big data processing by leveraging contemporary literature. The experimental benchmark verifies the applicability of the architecture. In our benchmark, we determine the performance of the small file problem on the distributed file system with a complex machine learning algorithm. The benchmark indicates that Lustre can substantially achieve better performance than HDFS, even though Lustre hardware resources, which do not leverage fast memory for cache, are very limited.

However, there are some limitations in our work. First, in terms of scalability, the architecture is only verified in a small-scale environment in which hardware resources do not meet the criteria of a production environment. Second, the amount of small files used for the benchmark is not big enough to absolutely reflect the consistent benchmark result of distributed file systems in extreme scenarios. Third, we did not include Tachyon in this proposed architecture to increase overall performance for both HDFS and Lustre file systems. Finally, we did not cover a write benchmark, a crucial factor for scientific computing and big data processing. 

In future work, we will verify the applicability of the proposed architecture in a larger-scale environment with a larger number of small files to be able to determine hidden problems that are impossible to find in a small-scale environment. In addition, we will also further investigate the impact of Tachyon on overall performance. Furthermore, by using \textit{libhdfs} (JNI based C API for Hadoop HDFS), we will perform a write benchmark with small files for both Luster and HDFS.

% or


% \appendices
% \blindtext

% \section*{Acknowledgment}


\ifCLASSOPTIONcaptionsoff
  \newpage
\fi


\bibliographystyle{IEEEtran}
\bibliography{IEEEabrv,BIB}

\end{document}